# The Bit-Generator and Time-Series Prediction

E. EISENSTEIN, I. KANTER, D. A. KESSLER
*Minerva Center and Department of Physics, Bar-Ilan University*
*Ramat-Gan 52900, Israel*
E-mail: eisenst@bk1.ph.biu.ac.il, kessler@dave.ph.biu.ac.il, ido@kanter.ph.biu.ac.il

and

W. KINZEL
*Institut für Theoretische Physik, Universität Würzburg*
*D-97074 Würzburg, Germany*
E-mail: kinzel@physik.uni-wuerzburg.d400.de

ABSTRACT

We study the dynamics of the Bit-Generator: a perceptron where in each time step the input units are shifted one bit to the right with the state of the leftmost input unit set equal to the output unit in the previous time step. The long-time behavior of the Bit-Generator consists of cycles whose typical period scales polynomially with the size of the network and whose spatial structure is periodic with a typical finite wave length. We investigate the problem of training one Bit-Generator to mimic another. The generalization error on a cycle is zero for a finite training set and global dynamical behaviors can also be learned in a finite time. Hence, a projection of a rule can be learned in a finite time.

In this paper, we study the dynamics of the Bit-Generator, [1] (BG) a perceptron whose input is constantly updated using the output of the network at previous times. The generalization features of such a network is fundamentally different than that of a network trained on a given set of random examples. It is essentially an exercise in the prediction of time series.[2] While the classic learning problem involves assimilating some finite set of essentially random data, prediction of time series involves exposure to a subset of an infinite, but *temporally correlated* data stream. Induction, or time series prediction, clearly constitutes a large fraction of the natural activity of the human neural network. New tools and methods of analysis are needed to explore this rich and important area.

In the prototypical task of prediction in classical neural networks, learning a rule from **random** examples,[3,4,5] the performance of the network is measured by the generalization error, $\epsilon_g$, the probability that the learning algorithm will predict wrongly the classification of a new random input after it has seen a certain number of random inputs and their associated target output as defined by the rule. The generalization error has been calculated analytically and numerically for feedforward architectures trained by various algorithms.[3] One of the remarkable conclusions from this body of work is that for realizable rules and continuous weights, the optimal generalization error only falls off asymptotically for large $\alpha$ as $\epsilon_g \propto 1/\alpha$. Here $\alpha = m/N$ denotes the ratio of the size of the training set, $m$, to the dimension of the input, $N$.[3,4,5] The generalization behavior of a BG, as we shall see, is very different, with the dominant behavior being learnable with essentially perfect accuracy.

The BG, the focus of our studies herein, is based on the prototype of the feedforward architectures, the perceptron.[6] It is important to note that the perceptron is a classifier with binary output values. Although most applications employ continuous outputs, statistical mechanics studies of learning and generalization have shown that binary classifiers provide useful insights into more complex networks.[3]

The perceptron is a single layer classifier consisting of $N$ binary inputs $S_i$ and one binary output $o$. The $\mu$'th output, $o^\mu$, depends on the the $\mu$'th input via the $N$ dimensional random weight vector $W$ following the rule

$$o^\mu = sgn[W \cdot S^\mu] \quad . \tag{1}$$

In the classic studies of learning, the inputs were chosen from some given distribution, whereas in the BG the next input is chosen to depend on the state of the network in the previous time step in the following way. In each time step the input units are shifted one bit to the right with the state of the leftmost input unit set equal to the state of the output unit in the previous time step. Mathematically the process can be summarized by

$$S_1^{\mu+1} = o^\mu \ ; \quad S_i^{\mu+1} = S_{i-1}^\mu \quad i = 2, \ ..., \ N \tag{2}$$

where $S_i^\mu$ is the $i$th input of the $\mu$th pattern.

Let us now mention two interesting interpretations of the dynamical process defined by eqs. (1)-(2). One way of thinking about this process is as a 'machine' which generates an infinite sequence of bits from one initial state, the first input of $N$ spins. The first input generates a bit, which is appended to the left of the spins. The machine then acts on the $N$ leftmost bits to generate another bit which is appended to the left, and so on. This is the origin of the name Bit-Generator devoted to this dynamical process.[7]

From a different point of view the dynamical process defined by the BG is equivalent to a sequential updating of a fully connected asymmetric network consisting of $N+1$ Ising spins, where the interactions are only a function of the difference between the 'location' of pairs of spins (assuming periodic boundary conditions)[8]

$$W_{ij} = G_{i-j \mod N+1} \quad . \tag{3}$$

The matrix $W$ is a Toeplitz matrix with $N$ independent interactions. $G_1, \ ..., \ G_N$, and zeroes on the diagonal. This mapping to a fully connected system may be interpreted as a form of data compression where each degree of freedom appears only once.

Many questions may be asked regarding the dynamical behavior of the BG. On one hand, the representation of the BG as a perceptron generating a sequence of bits raises the question of estimating its learning ability and generalization performance, especially in comparison to the generalization performance in learning from random examples. On the other hand, the mapping of the BG to a sequential updating of a fully connected asymmetric discrete spin system raises the question whether the long-time behavior of the system is governed by chaotic flows or stationary states. In the following it will become clear that the features of the BG in these two 'different' aspects are strongly coupled.

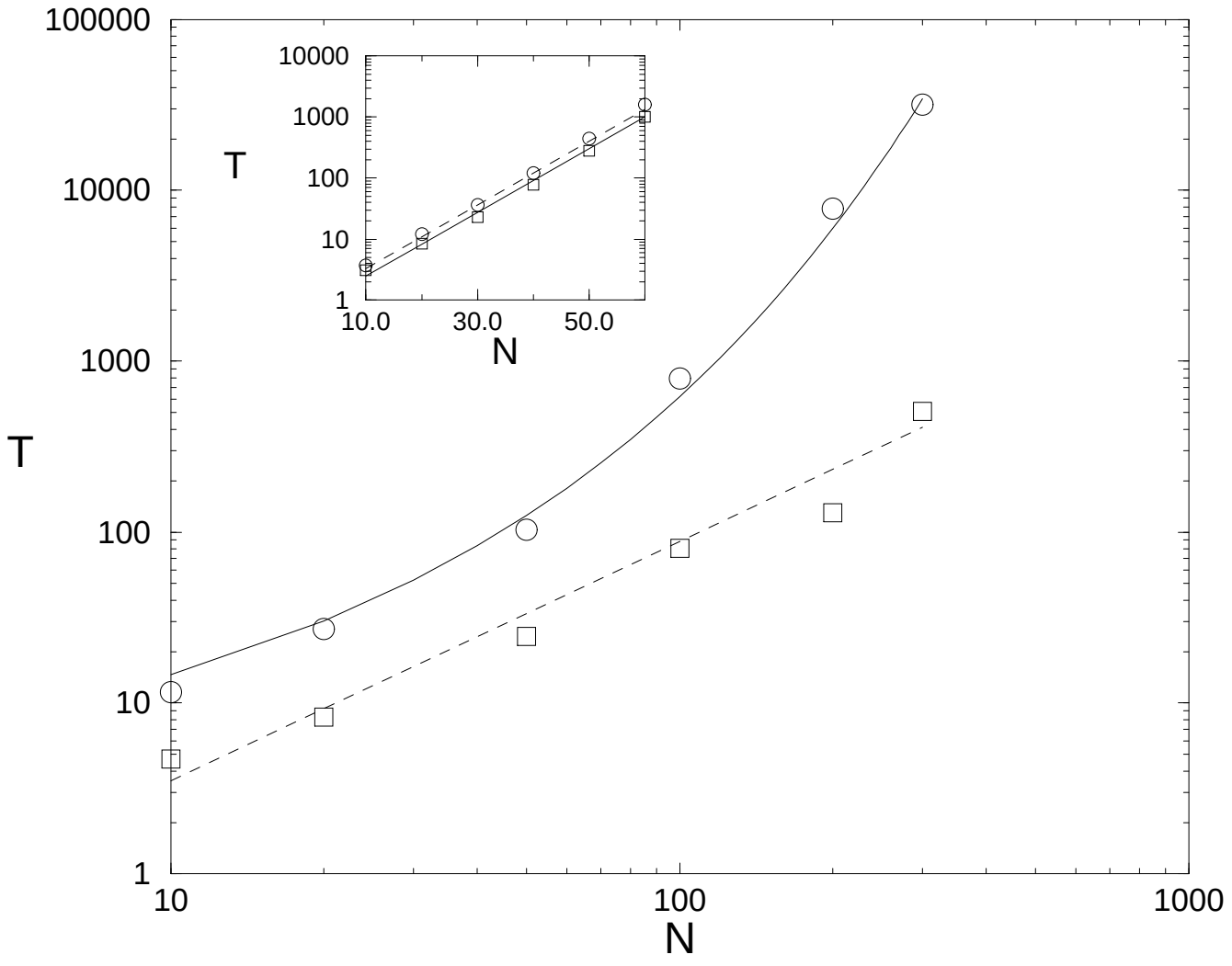

Fig. 1. The ensembled-averaged maximal period, $T_{max}$ (solid) and median period (dashed) for the BG. Inset: The same quantities for the uncorrelated case, ref. 8.

A related system consisting of $N$ Ising spins where each $W_{ij}$ is an independent random variable has been investigated analytically[9] and found to possess two remarkable features: (1) the system flows into cycles whose periods scale exponentially with the size of the system, (2) the correlation between two states of the system decays to zero after one Monte-Carlo step per spin. (These features are common to a broader class of models including the random map model.[10]) Note that the two main differences between this system and the BG are (1) the spatial structure of the matrix $W$ in the case of the BG and (2) the number of degrees of freedom of the weights scales differently in $N$, $O(N^2)$ and $O(N)$ respectively. Either of these differences is sufficient to prevent the mapping of the fully connected system to a BG. Surprisingly, it is found that only the spatial structure of the matrix $W$ affects dramatically the behavior of the BG. The averaged maximal period was found in the simulations to scale exponentially with $\sqrt{N}$ (see. Fig. 1). The average was performed over many realizations of the matrix $W$, where in each realization the maximal period was determined from a large number of different random initial conditions. This result is in contrast to the case of an unstructured matrix $W$ where $T_{max} \sim e^{0.12N}$ (see Fig. 1), independent of the number of degrees of freedom for the weights. Note that each step in these measurements stands either for one Monte-Carlo step per spin in the fully connected system or $N+1$ steps in the BG representation. (A fractional period was never encountered in our simulations on large systems.) Furthermore, the median period was found to scale polynomially with the system's size as $T_{med} \sim N^{0.8}$ (see Fig. 1). This is in contrast to the unstructured case and the random map model, where the avarage value of T scales as $e^{0.072N}$ and $e^{0.346N}$, respectively.[10] Note that the $T_{max}$ was found to be self-averaging,[11,12] since the simulations indicate that $< \ln(T_{max}) > \sim \ln(< T_{max} >)$, where $< ... >$ stands for an average over the samples. In the simulations each system was initiated from many random initial conditions, and proceeded to flow through a transient into a cycle. Each cycle was identified by the return of the BG to one of its previous states and its period measured.

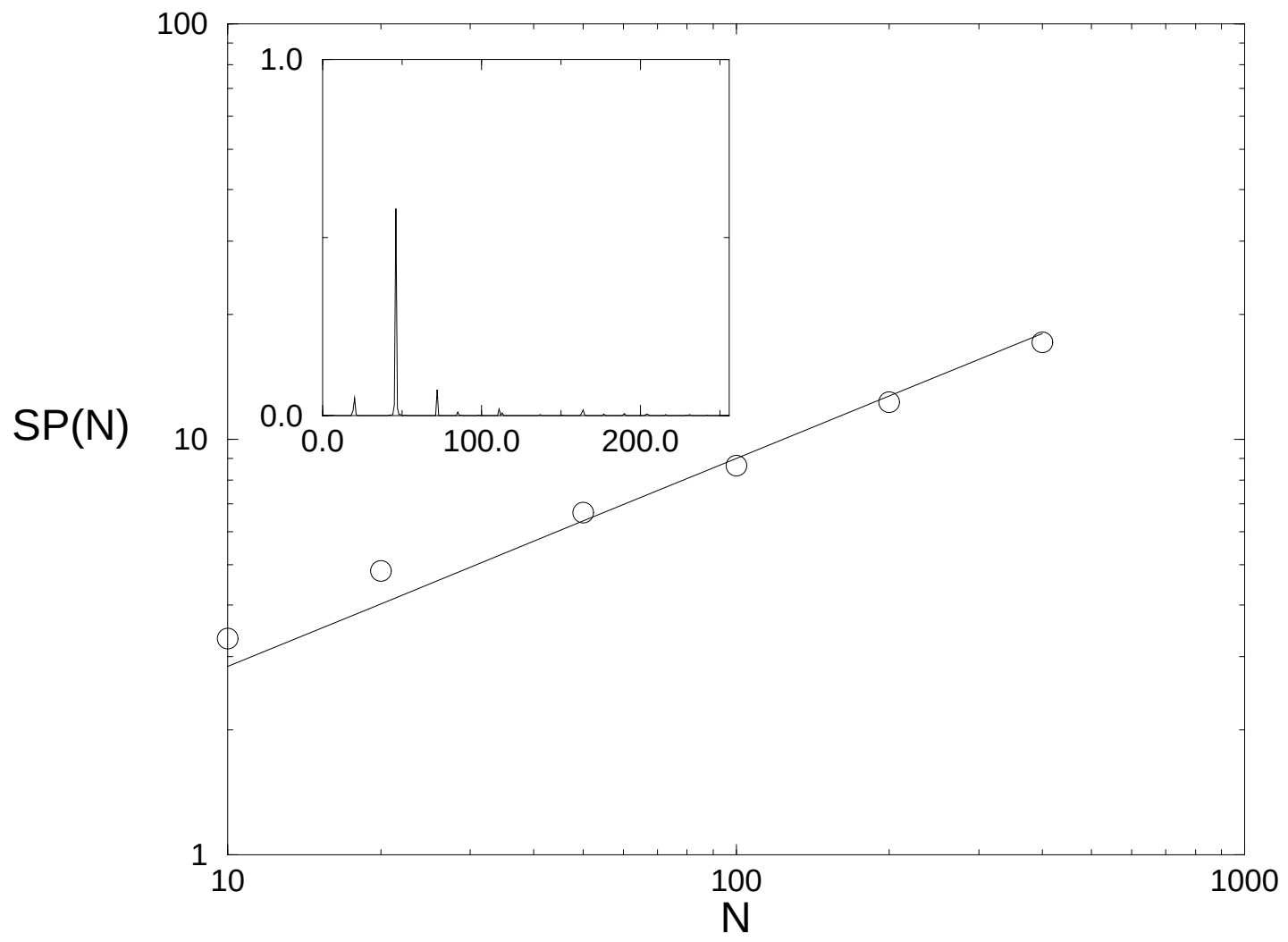

Fig. 2. The number of stable peaks,$SP$, as a function of $N$. Inset: Power spectrum of a state on a cycle for $N = 512$

The most striking result for the BG is that unlike the unstructured case where temporal correlations decay to zero at once, here the temporal correlations do not decay even asymptotically. A typical result for the overlap between a given input state on a cycle and subsequent input states is presented in Fig. 3. This mystery is resolved when the power spectrum of states belonging to a cycle is examined, displaying a sharp peak in the power spectrum associated with one of the $N/2$ possible wave numbers. An example of the power spectrum of a state belonging to a typical cycle is presented in the inset of Fig. 2. The height of this peak relative to the background is found to increase with $N$ with the peak consuming a constant fraction of the spectral weight as $N \to \infty$. This result indicates that all states belong to a cycle having a structure which is close to be a periodic spatial structure (such as $+-+-+-+-....$ for the wave number $k = N/2$, for instance). There are two main factors which prevent the power spectrum from being a delta function at one of the possible wave numbers. The first factor arises from the binary nature of the spins which is the reason for the observed higher harmonic terms with an appropriate weight. For instance, the first higher harmonic term appears at three times the basic harmonic term with a height $\sim 1/9$ that of the first harmonic. The second factor is the typically incomplete periodic structure. For illustration, assume that the dominant wave number is $k = N/2$ but there is a defect (boldface $+$ sign)in the structure $+ - + - + - \ldots - +\mathbf{+}+ - + - \ldots + - + - + -$. Note that even a finite small fraction of such defects does not affect the phase coherence and so does not destroy the dominant peak in the power spectrum. In the example of Fig. 3, one can see the periodicity is equal to 85, whereas the main peak corresponds to a wavelength of 17.

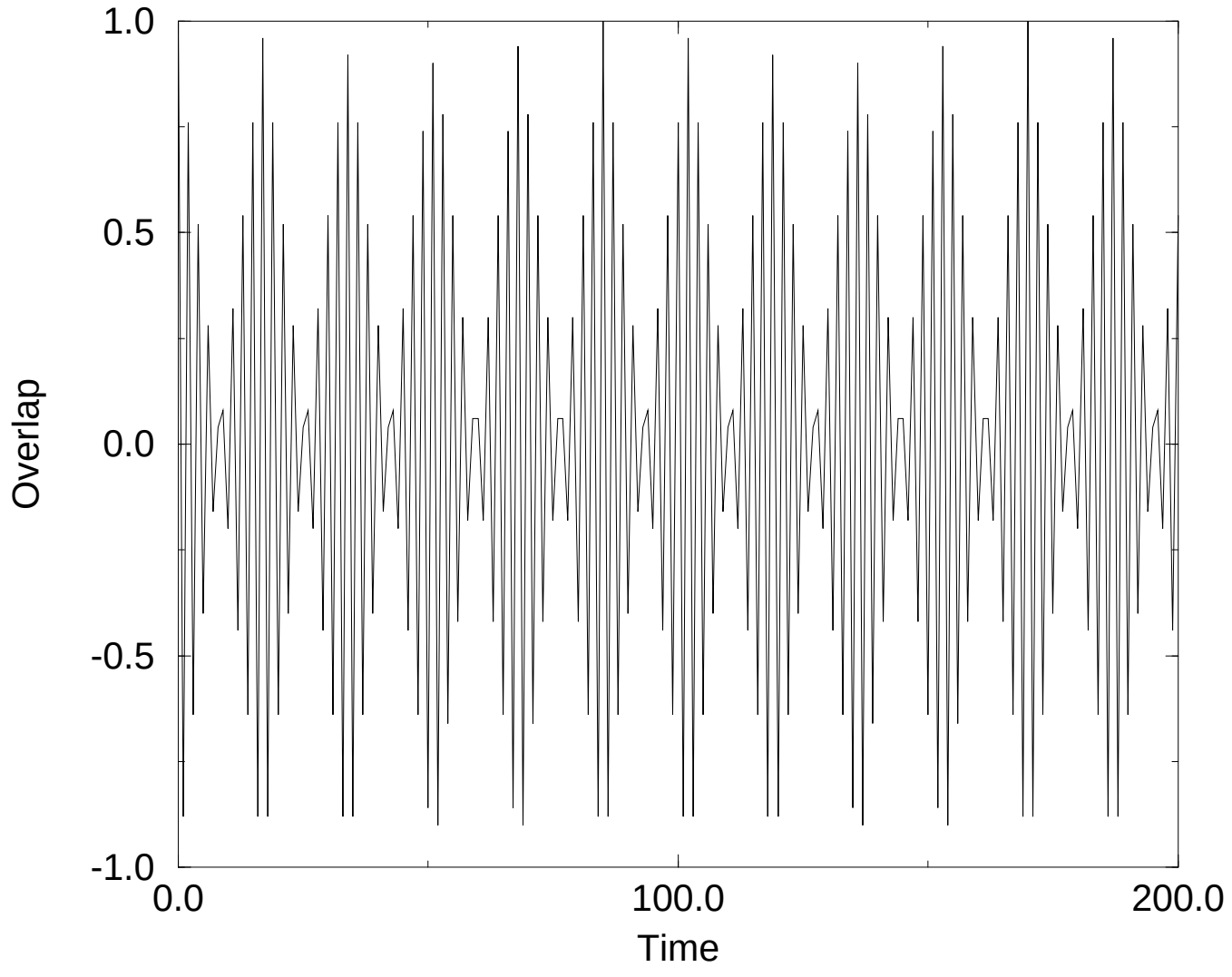

Fig. 3. The overlap between an input state on a cycle and subsequent input states as a function of time. $N = 100$

In the simulations it was found that many cycles have very small basins of attractions. Furthermore, in contrast to the behavior of the random map model[10], the size of the basin of attraction does not necessarily relate to the cycle's period, and usually the longest cycle has a very small basin of attraction. Hence, the examination of the nature of the cycles of the BG using an algorithm which flows into cycles from many random initial conditions is inefficient, especially for large system sizes where the length of the transient also becomes very large. However, using the knowledge that the power spectrum of a state on a cycle always has only one dominant peak suggests a much simpler algorithm to investigate the features of the cycles in the system. In this algorithm the initial conditions are states which are close to each one of the $N$ pure wave numbers. There are many ways to implement these initial conditions and, for instance, one can use $S_l = sgn[\cos[2\pi lk/N]]$ where $sgn(0)$ alternates sequentially between 1 and $-1$.

Using this idea that each cycle is associated with a single dominant wave-number (a peak in the power spectrum), the average number of such 'stable peaks' characterizing the cycles of a BG is calculated numerically. It is found that the number of stable peaks, averaged over realizations, scales with $\sqrt{N}$, see Fig. 2. An explanation of this scaling behavior and an exact relation between the power spectrum of the weights, $W$, and the wave numbers of the stable peaks is unclear. Nevertheless, from our simulations one can conclude the following heuristic rules regarding the question of what are the $\sqrt{N}$ stable peaks among the $N/2$ possibilities. (1) A stable peak is associated with one of the dominant peaks in the power spectrum of the weights. (2) Among these dominant peaks only some of the isolated peaks survive. An isolated peak means a dominant peak in the power spectrum of the weights which is surrounded by much smaller peaks. Using a similar algorithm, where many initial conditions associated with each one of the wave numbers but with a different structure to the defects were used to calculate the number of different cycles. It was found that there are typically a few different cycles associated with each stable peak, and therefore there are number of ways to embed the defects for each wave number.

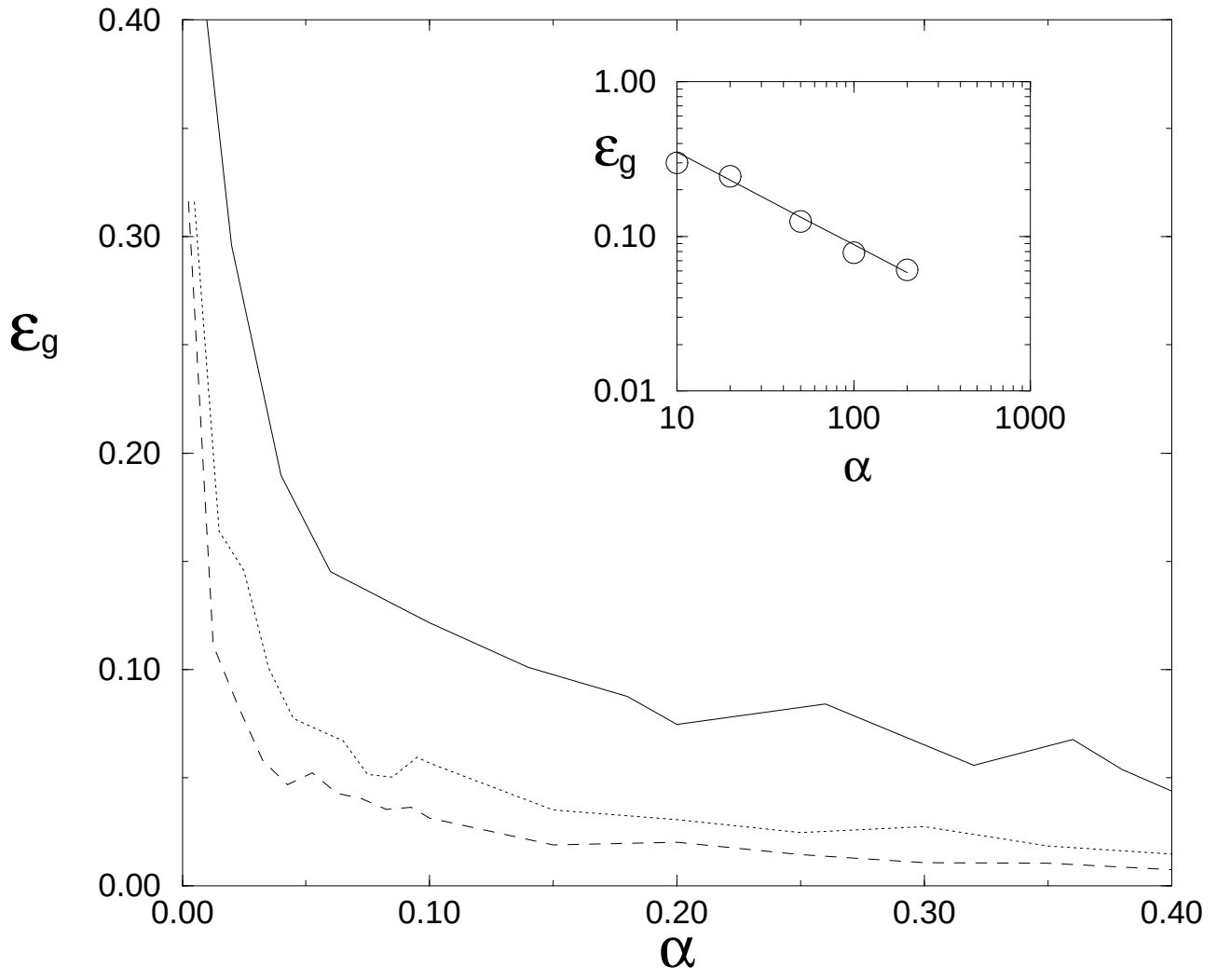

Fig. 4. The generalization error as a function of $\alpha$ for $N = 50$ (solid), 200 (dotted) and 400 (dashed). Insert: The generalization error as a function of $N$ for fixed $\alpha = 0.1$.

Let us now move from the investigation of the statistical nature of the trajectories in the phase space to examine the generalization performance of the BG. The definition of the generalization error, $\epsilon_g$, is the probability that a perceptron which has been trained to predict correctly a certain number of **consecutive** input/output pairs generated by a BG, predicts wrongly the next output. There are a few possibilities as to which sub-sequence to choose as a training set. The first case to be examined is where the training set is taken from a cycle produced by the dynamics of the BG from a random start. The result of the simulations for $\epsilon_g$ as a function of $\alpha$, where $\alpha N$ is the size of the training set, are presented in Fig. 4 for $N = 50,\ 200$ and 400. These results indicate the following: (1) at fixed size $N$ of the BG, $\epsilon_g$ decreases significantly even for small $\alpha$ and decays to zero at some $\alpha < 1$, (2) $\epsilon_g$ is a decreasing function of $N$. In order to find the asymptotic behavior as $N \to \infty$, $\epsilon_g$ was calculated for various $N$ at a fixed $\alpha$. The results for $\alpha = 0.1$, for instance, are presented in the inset of Fig. 4 and indicating that $\epsilon_g \propto 1/N$. Hence, in the thermodynamic limit $\epsilon_g$ is expected to be equal to zero for any finite $\alpha$. The explanation of this surprising result is based on the non-trivial spatial structure of the states belonging to a cycle. As discussed above, the power spectrum of such states are characterized by one dominant peak at a wave number $k$, indicating a periodic structure with periodicity $\lambda = N/k$. Hence, once the size of the training set is greater than or equal to $\lambda$, $\epsilon_g \sim 0$, if we may neglect the effect of the defects in the periodic structure. Since $\lambda < N$, it is clear that $\epsilon_g = 0$ for $\alpha > 1$. However it does not explain why $\epsilon_g = 0$ even for $\alpha < 1$. The explanation lies in the typically small spatial period of a state belonging to a cycle. This question was examined numerically, and it was found that $P(k)$, the probability that the dominant wave number of a cycle is $k$, is almost independent of $k$ (see Fig. 5). Hence, the typical wave number is of $O(N)$, the typical $\lambda$ is finite, and therefore $\epsilon_g = 0$ for any finite $\alpha$. The flat distribution of $P(k)$ also explains the result that $\epsilon_g \propto 1/\sqrt{N}$ for a fixed $\alpha$, since errors occur only for the $O(1/\sqrt{N})$ stable $k < 1/\alpha$. It is important to note that as a consequence of the periodic structure, the generalization error to predict the output far in the future is also essentially zero for all $\alpha > 0$.

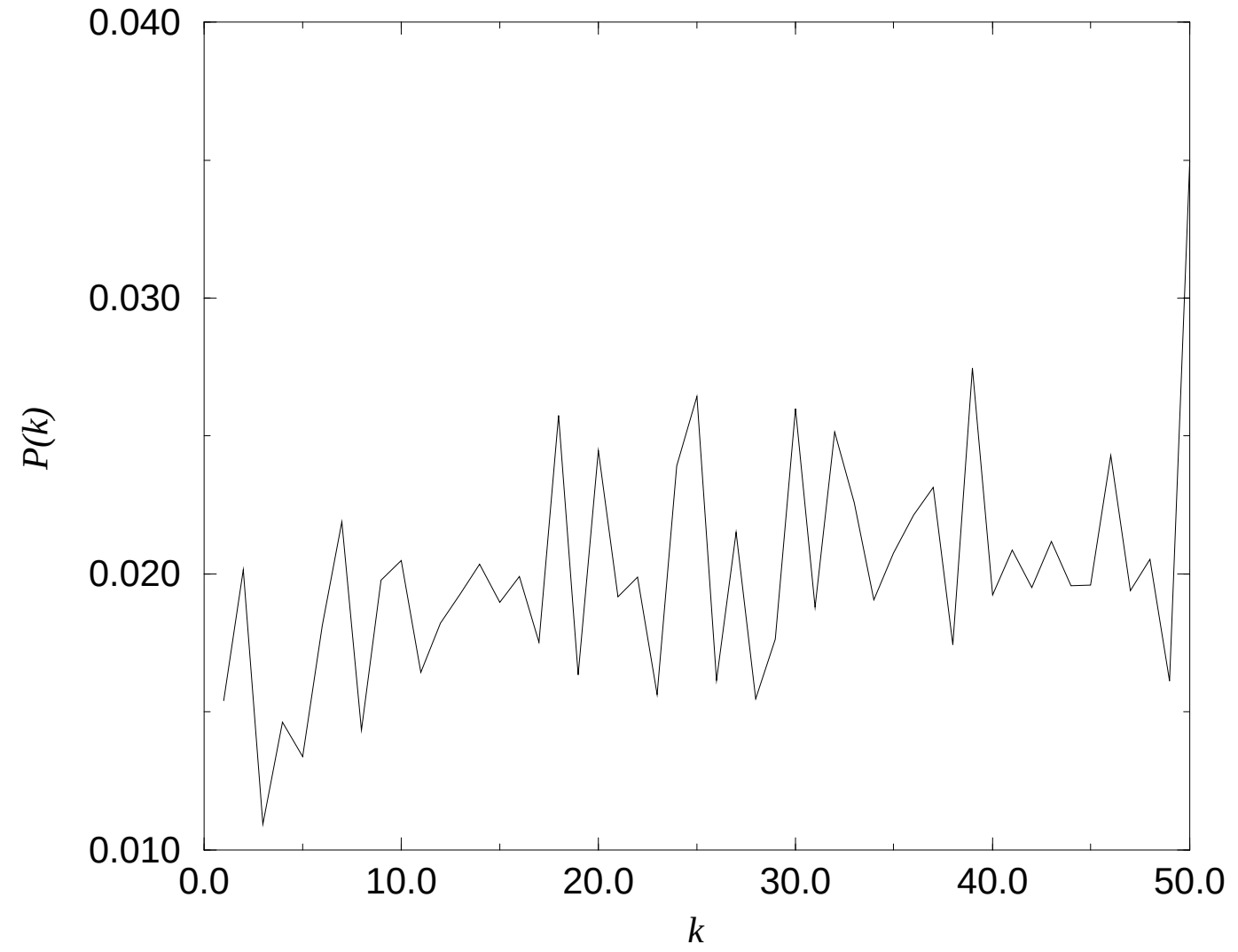

Fig. 5. The probability distribution for the dominant wavenumber $k$, $P(k)$ for $N = 100$, averaged over 1000 samples.

Another striking result is that a perceptron which has been trained to predict correctly $\alpha N$ consecutive input/output pairs which are taken from a periodic flow of the BG, learns also the period of this particular cycle, besides its spatial structure. The result for the probability distribution of the ratio between the temporal period of the cycle of the BG (the teacher), $T_t$, and that of the trained perceptron (the student), $T_s$, is presented in Fig. 6 for $\alpha = 0.1$ and 0.5 for $N = 100$. As $N$ increases, the minimum of $T_s/T_t$ occurs at a smaller $\alpha$ and it is plausible that as $N \to \infty$, $T_s/T_t \to 1$ for any finite $\alpha$. At $\alpha = 0$ there is no correlation between the two networks, and since, as was discussed above, the periods of the cycles fluctuate in leading order, the median period scales differently than the maximal one so that, for instance, $T_s/T_t$ diverges to infinity. However, at a finite $\alpha$ the trained network is capable at learning a **global dynamical quantity** - the size of the cycle. This is especially non-trivial given the multiplicity of different periods associated with each stable wavenumber.

The results that on a cycle $\epsilon_g = 0$ for any finite $\alpha$ and that even global dynamical quantities can be learned indicate that it is possible to learn perfectly in a finite time a projection of the features of a BG. Learning perfectly a rule of a network with continuous weights is impossible, however, the learning of a projection of a rule is possible to accomplish in a finite time.

The generalization performance of the BG during its flow in the transient to a cycle is beyond the scope of this paper, however we would like to briefly mention some preliminary results. The average length of the transient was measured and found to scale with $\ln N$, where each time step is equal to one step per spin in the fully connected network. This result can be well understood when the $N$ Ising spins are replaced by continuous spin variables with a spherical constraint, $\sum S_i^2 = N$.[8] Starting from random initial condition, the projection on the wave number $l$ of the Toeplitz matrix $W$ is $\propto \lambda_l^t / \sum_i \lambda_i$, where $t$ measures the number of steps. It is clear that the projection of $\lambda_{max}$ is the only dominant term when $t \propto \ln N$. The generalization error on the transient was also measured and found to be smaller in comparison to the generalization error of learning from random examples, since the examples in the BG case become correlated as one of the peaks becomes dominant.[13]

Preliminary results indicate that the maximal length of a random binary sequence that can be learned by a BG is $\sim 1.75N$, where $N$ is the size of the BG.[14] In this case correlations among the input/output pairs decreas the capacity of the network. As the correlations are reduced the capacity associated with random input/output is obtained.

Finally we mention a few interesting extensions currently under investigation. First is the study of a BG where the architecture of the network consists of a feed-forward network with a hidden layer. A second is an extension of the dynamical rules such that the next output is a function of the previous $l$ input/output states. It would be also interesting to extend this work to continuous time series, examining in particular the existance of chaotic flows and the scaling of the generalization error with the number of learning steps.

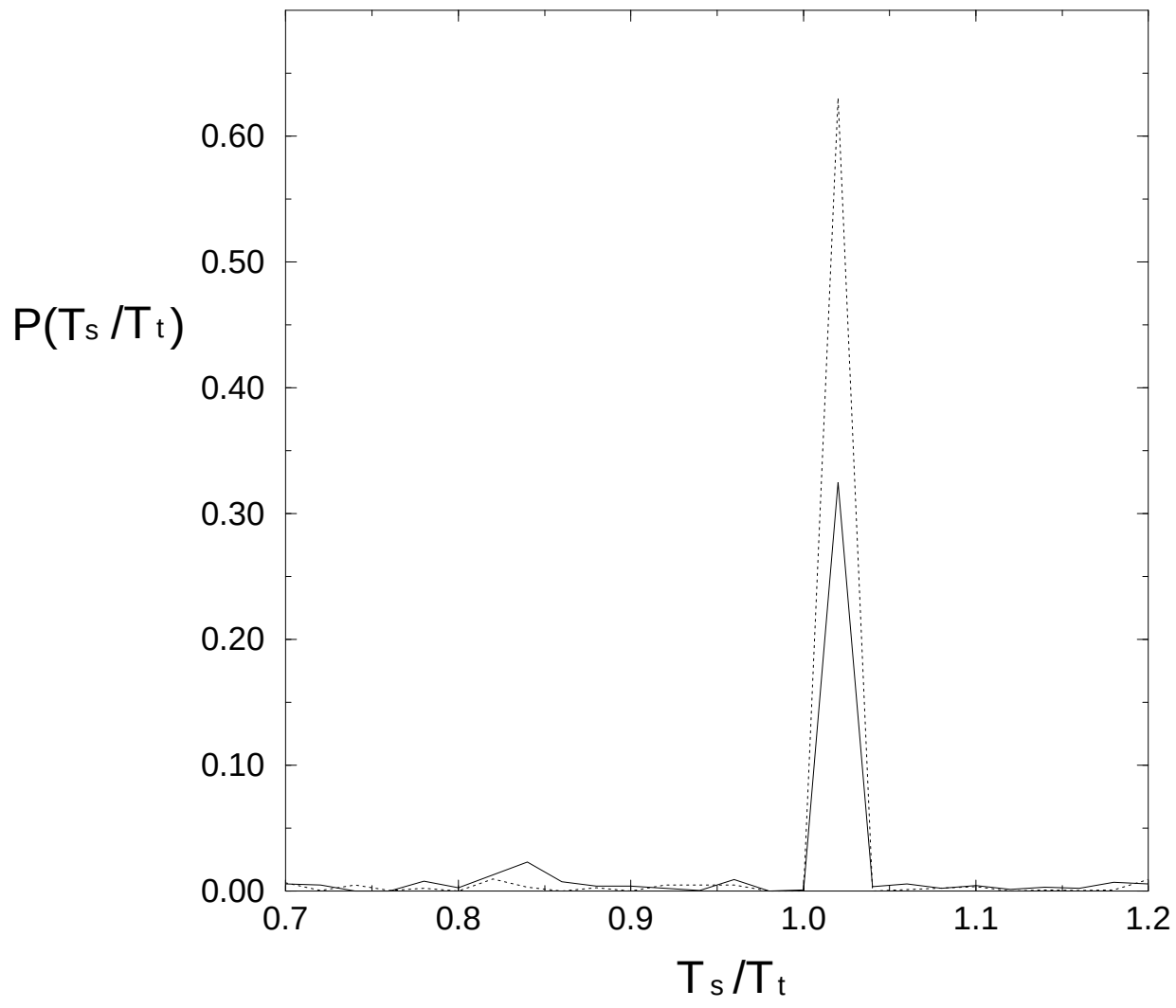

Fig. 6. The probability distribution $P(T_s/T_t)$ for $\alpha = 0.1$ (solid) and 0.5 (dotted) for $N = 100$.

## 1. Acknowledgements

DAK acknowledges the support of the Israeli Acadamy of Sciences and the Raschi Foundation. IK acknowledges the support of the Israeli Academy of Sciences.